\documentclass[11pt]{article}

\usepackage{epsfig,latexsym}
\newcommand{\sect}[1]{\setcounter{equation}{0}\section{#1}}

\def\aaa{angular momentum~}

\def\AAA{{\cal A}}

\def\ba{\begin{eqnarray}}

\def\bbb{background~}
\def\bbbb{backgrounds~}
\def\be{\begin{equation}}
\def\bi{\bibitem}

\def\Bar {\overline}
\def\BB{{\cal B}}

\def\de{\delta}

\def\di{\partial}
\def\De{\Delta}
\def\ea{\end{eqnarray}}
\def\ee{\end{equation}}
\def\eee{equation~}
\def\eeee{equations~}
\def\em{energy-momentum~}
\def\ep{\epsilon}
\def\eq{\equiv~}
\def \EE{{\cal E}}

\def\fr{\frac}
 
 \def\gmn{g_{\mu\nu}}
\def\Ga{\Gamma}

\def\ha{\frac{1}{2}~}

\def\in{\infty}

\def\II{{\cal I}}

\def\JJ{{\cal J}}
\def\ka{\kappa}

\def\kkk{Killing vector~}
\def\kkkk{Killing vectors~}
\def\la {\lambda}
\def\lll{\left(}

\def\LLL{\left[}

\def\mn{\mu\nu}

\def\na{\nabla}
\def\nl{\newline}
\def\nn{\nonumber}
\def\nnn{\noindent}

\def\om{\omega}

\def\Om {\Omega}

\def\ra{\rightarrow}

\def\rrr{\right)}

\def\RR{{\cal R}}

\def\RRR {\right]}
\def\si{\sigma}

\def\sss{spacetime~}
\def\ssss{spacetimes~}
\def\sup{superpotential~}

\def\Sc{Schwarzschild~}

\def\td{\tilde}
\def\te{\theta}
\def\th {\frac{1}{3}}

\def\vf{\varphi}

\def\vp{\varpi}

\def\vs{\vskip 0.5 cm}

\def\we{\wedge}

\def\ze{\zeta}

\def\1k{\fr{1}{\ka}}
\def\2k{\fr{1}{2\ka}}

\title{{\bf Gravitational Energy}}
\author{ Joseph Katz\thanks{email: jkatz@phys.huji.ac.il} 
\\{\it  The Racah Institute of Physics, Edmond Safra Campus}\\ {\it Givat Ram,
Jerusalem 91904, Israel}\\
\nl }

\begin{document}

\maketitle
\begin{abstract}

 Observers at rest in a stationary \sss flat at infinity can measure small
amounts of rest-mass+internal energies+{\it  kinetic energies}+pressure energy
in a small volume of fluid attached to a local  inertial frame. The
sum of these small amounts is the total ``matter energy" $E_M$ for those
observers. If $Mc^2$  is the total mass-energy, the difference
$ Mc^2-E_M$  is the binding gravitational energy. 

 Misner, Thorne and Wheeler evaluated  the   gravitational
energy of a spherically symmetric static spacetime. Here we show how to
calculate gravitational energy in any static and stationary \sss with
isolated sources with a set of observers at rest.

The result of MTW is recovered and we find that electromagnetic and
gravitational 3-covariant energy densities in conformastatic spacetimes are of opposite
signs. Various examples suggest that gravitational energy is negative in \ssss with
special symmetries or when the energy-momentum
tensor satisfies usual energy conditions.
\nl\nl
PACS numbers 04.20. -q, 04.20.Cv
	 
\end{abstract}
 
\sect{Introduction }
 
 The paper deals  mainly with stationary \ssss
of isolated sources that are asymptotically flat. 

In classical physics, energy has different forms which are   additive. If energy is
conserved one can assess how much change  and transfers of energy occurs
between the various forms. In Einstein's theory of gravitation different forms of
matter energy are mixed in inseparable ways with   gravitational binding energy.
Gravitational energy is diffuse, partly mixed up with other forms, with matter
energy, partly stored in the gravitational field itself. Alas, the matter is the
source of gravity  so that   it is impossible to disentangle gravitational energy
from other forms of energy  with or without  solving the field equations. This is a
great loss  that adds to that of ``gravitational  force" and   ``gravitational
energy density" with which it shares a common origin: the principle of
equivalence.

One may, however, define  gravitational energy with respect to a set of
observers. Take for instance  the matter-energy tensor
     of a perfect fluid; in standard notations\footnote{Indices $\la,
\mu, \nu, \rho,\cdots=0,1,2,3 $; indices
$k, l, m, n\cdots=1,2,3$, the metric
$\gmn$ has signature $+---$ and $g$ is its determinant. Multiplication by
$\sqrt{-g}$ is indicated by an additional hat like $\hat X$ on a previously defined
symbol like $X$.  Covariant derivatives are represented by a $D$ and partial
derivatives by a $\di$. And for once neither $G$ nor $c$ are set equal to
$1$ and
$x^0\equiv ct$. Finally the Levi-Civita symbol is $\ep_{\mu\nu\rho\si}$ with
$\ep_{0123}=1$.}   $  T^\mu_\nu=(\rho c^2+P)u^\mu u_\nu-\de^\mu_\nu
P$ and let $w^\mu$ be  the 4-velocity field of   local observers ($\gmn w^\mu
w^\nu=1$). If
$dV_\mu=\fr{1}{3!}\ep_{\mu\nu\rho\si} dx^\nu \we   dx^\rho\we  dx^\si$ is
the 3-dimensional  coordinate  volume element of a spacelike hypersurface [say, for
instance, 
$d^3x$ on
$t=0$ in adapted coordinates $(x^0=ct,x^k)$],   the scalar element
\be 
dE_M\equiv \hat T^\mu_\nu w^\nu dV_\mu=\fr{d(M_0c^2)}{\sqrt{1-\vec
v^2}}+dE_I~~~,~~~d(M_0c^2)=\rho c^2\sqrt{-g}u^\mu dV_\mu~~~{\rm
and}~~~u_\nu w^\nu=\fr{1}{\sqrt{1-\vec v^2}}.
\ee 
  $d(M_0c^2)$ is the element of rest mass + internal energy of the
source, $\vec v$ the velocity of the matter with respect to the local observer in
a local inertial frame  and $dE_I$ is the external potential energy of the
pressure on the small proper volume element $\sqrt{-g}u^\mu dV_\mu$. 

One  
``natural" selection of observers in   static
\ssss which have a field of timelike
\kkkk of translations
$\xi^\mu$ are observers with velocities
$w^\mu=\xi^\mu/\xi$. They are at rest in any system of coordinates that singles
out the inertial time   at infinity
$\xi^\mu=\{1,0,0,0\}$.  They are also at rest with respect to the
matter. There is, however, no such        well defined set of observers in
stationary spacetimes. For instance, in a Kerr \sss one  might    prefer to
choose the family of ``zero
\aaa observers", the ZAMOS  of Bardeen
\cite{Ba}. It is of course desirable to have some unique way of choosing
observers in stationary spaces. Such a qualifying choice is not considered in
this work.

 The sum
$E_M$ of
$dE_M$'s   can be regarded as the total``  rest mass-energy+internal
energies+kinetic energy+ pressure energy" to which it  reduces   in the weak
field limit. No gravitational energy is involved in $dE_M$.  Therefore if
$Mc^2$ is the total mass-energy of spacetime the difference
$E_G\equiv Mc^2-E_M$  can   be regarded as the total gravitational energy
 of  spacetime for our set of  observers. When gravity is a binding force, and
this is not necessarily the case in general relativity, one expects that like  in
Newtonian gravity  
$E_G$ is negative.  

Misner, Thorne and Wheeler \cite{MTW} evaluated $E_G$ along
precisely this line of taught for  
\ssss created by  static spherical ``stars". Here we  extend the
 calculation of   
$E_G$ to any  static spacetime with localized sources. The formula holds also
in
 stationary spacetimes but its application may generate less enthusiasm
because, as we said, sets of reference observers are not as well defined.   In
the few examples we treat
$E_G<0 $  when \ssss have special symmetries or when the \em tensors of the source
satisfy usual energy conditions.


 \sect{Conservation laws and   observers}
 
 {\it (i) The   conserved current for energy}
\vs 
To begin with we need a conserved covariant vector  $J^\mu$ involving
the source of gravity
$T^\mu_\nu$. There are a number of
available vectors, see in particular    Abbott and
Deser
\cite{AD},   Grishchuk, Petrov and Popova \cite{GPP} or  \cite{KBL}\footnote{
The Ashtekar and Hansen
\cite{AH} conformal mass formula, the Hamiltonian formulation  of Regge and
Teitelboim
\cite{RT} and various formulas assembled by Szabados \cite{Sz} to calculate energy in
a finite volume (quasi-local energy)  are integrals on closed surfaces. The integrant is
an antisymmetric tensor whose diverge is a conserved vector which must be of the
same form as eq. (\ref{21}). One may thus start from   those formulas as well  but this
needs a bit of additional work.}. All conserved
currents for energy have the same form in stationary spacetimes:
\be 
J^\mu=(T^\mu_\nu+t^\mu_{~\nu})\xi^\nu.
\label{21}
\ee 
 $t^\mu_{~\nu}$, which depends   on the metric and its derivatives, is
associated with part\footnote{It is well known, see for instance  \cite{LK}, that
the integral of
$\hat T^\mu_\nu\xi^\nu dV_\mu$ in the weak field limit   contains twice the
Newtonian gravitational energy.} of G-energy.
$t^\mu_{~\nu}$   differs from author to author   reflecting the ambiguity in
defining G-energy {\it density}.  

One necessary ingredient of $J^\mu$ is a background metric. In asymptotically
flat \ssss it is advantageous to introduce a flat
background. The background is   an extension from 
infinity inwards. It is   an artifact to get a covariant description and is very
useful when we want to use for instance spherical coordinates instead of
Minkowski coordinates. Mappings on the background  may be chosen at our
convenience.   The timelike
\kkkk associated with conserved energy are translations {\it in the background},
i.e.
\be
\Bar D_\mu\xi^\la=0\label{22}
\ee 
 An overbar refers to covariant derivatives in the background.

All covariant conserved vectors are also divergences of antisymmetric tensors
or superpotentials $J^{\mu\nu}$: 
\be
\hat J^\mu=\di_\nu\hat J^{\mu\nu}
\label{23}
\ee

The superpotentials are not   uniquely defined nor are they the same from author
to author. However, the  surface integral on a sphere at infinity of any
valid \sup must be the same i.e.
$Mc^2$.  
 
We shall here adopt\footnote{The \sup given in \cite{AD} does not   provide   
the Bondi mass at null infinity \cite{PK}  and the $t^\mu_{~\nu}$ of \cite{GPP}
 also given in \cite{PK}    contains second order derivatives of the metric which
is not a priori favorable to analyze the sign of the integrant.} the current given in
\cite{Ka} and more fully described in
\cite{KBL}. There is a uniqueness \cite{CN}\cite{JS} to the ``KBL-superpotential"
which has been found satisfactory in all applications at spatial as well as at null
infinity, on flat or non flat backgrounds including in the calculation of the very
exotic   mass of  Kerr black holes on   anti-de Sitter
\bbbb in $D$-dimensions \cite{DK}. And it is quadratic in first order derivatives of
the metric which   may be a real  boon as we shall see. In  \cite{KBL} 
  $t^\mu_{~\nu}$ is  the covariant form of the quadratic Einstein pseudotensor.

\vs  
 {\it (ii) The KBL conserved current}
\vs  
 Formulations in \cite{AD},   
\cite{GPP} and 
\cite{KBL} are all on generally curved backgrounds. We shall start
from the KBL identity $\hat I^\mu=\di_\nu\hat J^{\mn}$  which holds  not only
for curved
\bbbb but also for arbitrary vector fields which are still denoted by $\xi^\mu$ but
are not necessarily Killing vectors. Our $\hat J^\mu$ in (\ref{23}) is a
particular value of $\hat I^\mu$.  We shall first lay out the mathematical
expression of
$I^\mu$ and
$J^{\mu\nu}$, which are somewhat complicated, and remind their physical
content next. $\hat I^\mu=\di_\nu\hat J^{\mu\nu}$ is originally a N\oe ther 
identity which turns into a real conservation law  once Einstein's equations
$G^\mu_\nu=\ka T^\mu_\nu$ are taken into account, $ \ka=\fr{8\pi G}{c^4}$.  

The KBL current and
\sup were derived 
 in terms of Christoffel symbols of both the \sss  $\Ga^\la_{\mu\nu}$  and the
background $\Bar\Ga^\la_{\mu\nu}$ or, more precisely, in terms of a tensor
$\De^\la_{\mu\nu}\eq\Ga^\la_{\mu\nu}- \Bar\Ga^\la_{\mu\nu}$.  We found it
somewhat simpler to write $\hat J^\mu=\di_\nu\hat J^{\mu\nu}$ in terms of 
\be
 L_\rho^{~\mu\nu}\equiv \fr{1}{\sqrt{-g}}\Bar D_\rho \hat
g^{\mu\nu}~~,~~\De_\rho\equiv \ha\gmn L_\rho^{~\mu\nu}=\Bar
D_\rho\log \sqrt{-g}~~~{\rm
and}~~~\hat F^\mu\equiv \hat L_\nu^{~\mu\nu}=\Bar D_\nu\hat
g^{\mu\nu}.
\label{24}
\ee
  $\De^\la_{\mu\nu}$ and $ L_\rho^{~\mu\nu}$are so related:
\be
\De^\la_{\mu\nu}=\de^\mu_{(\rho}\De_{\si)}-L_{(\rho\si)}^{~~~~\mu}+\ha \lll
L^\mu_{~\rho\si}  -g_{\rho\si} \De^\mu \rrr.
\ee
The KBL \sup  which, in its original form is
\be
\hat J^{\mn}=\1k D^{[\mu}\hat\xi^{\nu]}+\1k \xi^{[\mu}\hat k^{\nu]}~~~ {\rm
with}~~~\hat k^ \nu =\fr{1}{\sqrt{-g}}\Bar D_\mu(-g\hat g^{\mn} ),
\ee
 can be rewritten as
follows\footnote{Our interest in the superpotential  here is limited. For a
detailed analysis of its structure and properties the reader is referred to the
original paper.}:
\be
\hat J^{\mu\nu}=\1k\lll \hat l^{\rho[\mu}\Bar D_\rho \xi^{\nu]} - \hat
L^{[\mu\nu]}_{~~~~\la}
\xi^\la+ 
\xi^{[\mu}\hat F^{\nu]}\rrr ~~{\rm where}~~\hat l^{\mu\nu}\equiv\hat
g^{\mu\nu}-\Bar{\hat g^{\mu\nu}}.
\label{25}
\ee
The current density $\hat I^\mu$ is linear in $\xi^\mu$, $\Bar D_\rho\xi^\mu $
and covariant derivatives of\footnote{$Z_{\rho\si}$ in \cite{KBL} is
here $2
\Bar z_{\rho\Bar\si}$. Also   indices in \cite{KBL} are   displaced with
$\Bar
\gmn$, never with $\gmn$. Here we do not stick to that convention. If an indice
is   displaced with the metric of the \bbb
$\Bar
\gmn$ the indice has an overhead bar like in $X_{\Bar\mu}=\Bar\gmn X^\nu$.
Otherwise  we   write $X_{ \mu}= \gmn X^\nu$.}
\be
\Bar z_{\rho\Bar\si}\equiv \ha (\Bar D_\rho \xi_{\bar\si}+\Bar D_\si
\xi_{\bar\rho})~~~{\rm in~which}~~~\xi_{\Bar\rho}\equiv \Bar
g_{\rho\la}\xi^\la
\ee
$\Bar z_{\rho\Bar\si}=0$ if $\xi^\mu$ is a \kkk of the background.
Thus\footnote{In its original form, see \cite{KBL}, the last two terms in (\ref{27})
are grouped slightly differently;   in the present notations we would have written
it like this
$\hat\si^{\mu[\rho\Bar{\si}]}\di_{[\rho}\xi_{\Bar{\si}]}+\hat\si^{\mu(\rho\Bar{\si})}\Bar
z_{\rho\Bar{\si}}+\hat
q^{\mu\nu\rho\si}\Bar D_\nu\Bar {z}_{\rho
\Bar\si}\eq\hat\si^{\mu[\rho\Bar{\si}]}\di_{[\rho}\xi_{\Bar{\si}]}+\hat\ze^\mu$
and $\hat\ze^\mu$ is given by eq. (2.36) in \cite{KBL}.}. 

\be
\hat I^\mu=\lll \hat T^\mu_\nu-\Bar { \hat T^\mu_\nu 
}\rrr\xi^\nu+\lll\fr{1}{2\ka}\hat l^{\rho\si}\Bar R_{\rho\si}\de^\mu_\nu+\hat
t^\mu_{~\nu}\rrr\xi^\nu+\hat\si^{\mu\nu}_{~~~\la}\Bar{D_\nu}\xi^\la+\hat
q^{\mu\nu\rho\si}\Bar D_\nu\Bar {z}_{\rho \Bar\si}.
\label{27}
\ee
The three undefined quantities in this vector are
\ba
t^\mu_{~\nu}&\equiv&S^\mu_{~\nu}-\ha\de^\mu_{~\nu}S ~~~{\rm
with}~~~S^\mu_{~\nu}\equiv\2k\LLL\lll  \ha L^{\mu\rho\si}  -L^{\rho\si\mu}
\rrr L_{\nu\rho\si}-\De^\mu\De_\nu\RRR,\\
\si^{\mu\nu}_{~~~\la}&\equiv&\2k \lll
L_\la^{~~\mu\nu}-F^\mu\de^\nu_\la-2L^{[\mu\nu]}_{~~~~\la} \rrr~~, ~~  
\hat q^{\mu\nu\rho\si}= \2k \lll  \hat l^{\mu\nu}  \Bar g^{\rho\si}+\hat
l^{\rho\si}  \Bar g^{\mu\nu} -2\Bar g^{\mu(\rho}\hat l^{\si)\nu}\rrr.\nonumber\\
 \ea
The most significant properties of the conserved vector (\ref{27})   were
pointed out in \cite {KBL}: (1) On a flat background ($\Bar
R^\la_{~\nu\rho\si}=0$)   with a timelike Killing vector of translations,
(\ref{22}),    
\eee (\ref{27})   reduces to (\ref{21}) where $\hat J^\mu$ is the special  
value of $\hat I^\mu$ in this case. The presence in the right hand side  of 
$\hat T^\mu_\nu \xi^\nu$ shows
that we are dealing with   conservation   of energy.  The total energy is   given by a
surface integral of the
\sup (\ref{25}).  (2) If the \kkk is one of rotations the surface integral gives the
corresponding total angular momentum components.
(3) On other \bbbb with various symmetries we obtain other conserved
quantities. 

A remarkable feature of the conserved
  vector is that  it is   not useful for the very reason
indicated in the introduction. There are no coordinate independent and background
independent properties which one can associate with the different terms of
$J^\mu$.  This will, however, change if we introduce a preferred set of local
observers.

\vs
 {\it (ii) Singling out the matter energy vector}
\vs
What we really want to see in $J^\mu$ is $T^\mu_\nu w^\nu$ rather than
$T^\mu_\nu
\xi^\nu$. 
We thus rewrite the first term of (\ref{27}) like this:
\be
 \lll \hat T^\mu_\nu-\Bar { \hat T^\mu_\nu 
}\rrr\xi^\nu=\lll \hat T^\mu_\nu-\Bar { \hat T^\mu_\nu 
}\rrr w^\nu+\lll \hat T^\mu_\nu-\Bar { \hat T^\mu_\nu 
}\rrr\tau^\nu  ~~~{\rm where}~~~\tau^\nu\equiv
\xi^\nu-w^\nu.
\label{211}
\ee
Ultimately $w^\nu$ will be a field of timelike unit vectors and $\xi^\nu$ a
timelike \kkk field but for the moment both are unspecified to leave
open the possibility  to use the formulas in non stationary \ssss and  
\ssss with non-flat backgrounds. We then replace
$\lll \hat T^\mu_\nu-\Bar { \hat T^\mu_\nu  }\rrr\tau^\nu$
in (\ref{211}) by its expression in terms of the Ricci tensor $R_{\rho \si }$ using Einstein's
\eeee in the foreground as well as in the background. Thus we set respectively
 \be
T^\mu_\nu \tau^\nu=\1k \LLL g^{\mu(\rho}\tau^{\si)}-\ha g^{\rho\si}
\tau^\mu
\RRR R_{\rho\si}\equiv \vp^{\mu\rho\si} R_{\rho\si} ~~~{\rm and}~~~ 
\Bar T^\mu_\nu \tau^\nu\equiv \Bar\vp^{\mu\rho\si} \Bar R_{\rho\si}.
\label{212}
\ee
A bar over the $\vp$'s indicates that the $g$'s have been replaced by $\Bar g$'s.
Next we use the expression of $R_{\rho\si}$   in terms of $\Bar
D$-covariant derivatives and $\De^\la_{\rho\si}$ as in \cite{KBL}\footnote{The
expression  (\ref{213}) given in
\cite{KBL} and to which we refer, for the facility of the reader since it is on the
web, was, however, published in 1940  by Rosen \cite{Ro}.} :
\be
R_{\rho\si}=\Bar D_\la \De^\la_{\rho\si}-\Bar D_\rho
\De^\la_{\si\la}+\De_{\rho\si}^\la\De_{\la\eta}^\eta-\De_{\rho\la}^\eta
\De_{\eta\si}^\la+ \Bar R_{\rho\si}. 
\label{213}
\ee
We replace   the $(T-\bar T)\tau$'s in (\ref{211}) by (\ref{212}) using 
(\ref{213}), change the $\hat\vp( \Bar D\De)$-terms into  [$\Bar D(\hat\vp
\De)-(\Bar D\hat\vp)\De$]-terms and obtain after  straightforward but
  slightly tedious calculations the following expression for   a  modified
conserved current  $\hat \II^\mu$ and superpotential $ \hat
\JJ^{\mu\nu}$ defined as follows:
\be
\hat\JJ^{\mu\nu}\equiv\hat J^{\mu\nu}-\1k \lll  
-\hat L^{[\mu\nu]}_{~~~\la}\tau^\la+ \tau^{[\mu} \hat F^{\nu]}\rrr=\1k\lll \hat
l^{\rho[\mu}\Bar D_\rho \xi^{\nu]} - \hat L^{[\mu\nu]}_{~~~~\la}
w^\la+ 
w^{[\mu}\hat F^{\nu]}\rrr.
\label{214}
\ee
Notice that $\hat\JJ^{\mu\nu}$ is similar to $\hat J^{\mu\nu}$; the $\xi$'s have
been replaced by $w$'s in the last two terms but not in the first one. 
$\hat \II^\mu$ is very similar to $\hat I^\mu$ with different coefficients:
\be
 \hat \II^\mu=\lll \hat T^\mu_\nu-\Bar { \hat
T^\mu_\nu  }\rrr w^\nu+ \hat\RR^\mu+ \lll   \hat t^\mu_{~\nu}w^\nu+\hat
t^{\mu\nu}_{~~~\la}\Bar D_\nu w^\la \rrr +
 \2k\lll \hat L_\la^{\mu\nu}-\de^\mu_\la\hat F^ \nu  \rrr\Bar D_\nu\xi^\la+\hat
q^{\mu\nu\rho\si}\Bar D_\nu\Bar z_{\rho\Bar\si},
\label{215}
\ee
in this
\be
\hat t^{\mu\nu}_{~~~\la}  \equiv \1k \lll \de^{[\mu}_\la \hat F^{\nu]} -
\hat L^{[\mu\nu]}_{~~~~\la}\rrr~~~{\rm and}~~~\label{216} 
 \hat\RR^\mu \equiv \2k\lll   \hat l^{\rho\si}\Bar R_{\rho\si}w^\mu+ \hat
l^{\mu\rho}\Bar R_{\rho\si}\tau^\si-  \hat l^{\rho\si}\Bar
R^\mu_{~~\rho\si\la}\tau^\la 
\rrr.
\label{217}
\ee
 
 {\it (iii) Gravitational energy in stationary spacetimes}
\vs
 The very complicated current $ \hat \II^\mu$ in (\ref{215}) with its 
\sup $ \hat
\II^{\mu\nu}$ in (\ref{214}) is considerably simpler on a flat background,
$\hat\RR^\mu =0$, with a timelike \kkk of translations, $\Bar z_{\rho\Bar\si}=0$ and
$\Bar D_\nu\xi^\la=0$; we denote it then by $ \hat\JJ^ \mu $:

\be
\hat\JJ^{\mu\nu}=\1k\lll   - \hat L^{[\mu\nu]}_{~~~~\la}
w^\la+ 
w^{[\mu}\hat F^{\nu]}\rrr\label{220}~~~{\rm and}  ~~~
\hat\JJ^\mu\eq\di_\nu\hat\JJ^{\mu\nu}=\hat T^\mu_\nu w^\nu+\lll   \hat
t^\mu_{~\nu}w^\nu+\hat t^{\mu\nu}_{~~~\la}\Bar D_\nu w^\la\rrr.
\label{221}
\ee
This new conserved current has the following properties. Let  $w^\mu$ be a
field of observers at rest, i.e.
\be
w^\mu=\fr{\xi^\mu}{\xi} ~~~{\rm
with}~~~{w^\mu}  {\longrightarrow} \xi^\mu+{\cal
O}^\mu(\fr{1}{r})~~~{\rm for}~~~r\ra\infty,
\label{218}
\ee
where $r\ra \infty$  is the radius of the infinite sphere  in the background
whose metric in spherical coordinates is
\be
\Bar{ ds}^2=c^2dt^2-dr^2-r^2(d\te^2+\sin^2\te
d\vf^2)\equiv c^2dt^2-dr^2-r^2d\om^2.
\ee
If $w^\mu$ is
replaced by the
\kkk $\xi^\mu$ in  $\hat\JJ^{\mu\nu}$ it is equal to the KBL
superpotential. Therefore the surface integral   of $   
\hat\JJ^{\mu\nu}$ at infinity is equal to $Mc^2$ because of (\ref{218}):
\be
\oint_{r\ra\infty}\ha\hat \JJ^{\mu\nu}dS_{\mu\nu}=
 \1k\oint_{r\ra\infty}\ha   
 \lll   - \hat L^{[\mu\nu]}_{~~~~\la}
\xi^\la+
\xi^{[\mu}\hat F^{\nu]}\rrr    dS_{\mu\nu}
=Mc^2~~,~~dS_{\mu\nu}=\ha\ep_{\mu\nu\rho\si}dx^\rho\we  dx^\si.
\ee
But the matter-energy
\be
E_M=\int_\infty \hat T^\mu_\nu w^\nu dV_\mu.
\ee
Therefore in a stationary spacetime the gravitational field energy
with respect to our set of observers  is given by an integral whose
integrant is quadratic in first order derivatives of the field:
\be
E_G=Mc^2-E_M=\int_\infty \lll   \hat t^\mu_{~\nu}w^\nu+\hat
t^{\mu\nu}_{~~\la}\Bar D_\nu w^\la \rrr dV_\mu.
\ee
This is an integral that gives the total gravitational energy in a stationary
\sss with localized sources. The integrant is quadratic in first order derivatives
and is covariant.  

 Now we examine  the gravitational energy in some particular static and stationary
spacetimes. 
\sect{Spherically symmetric static spacetimes}

 Considered in isotropic coordinates, 
the metric is
\be
ds^2= a^2(d{x^0})^2-b^2d\vec r{~^2}~~~{\rm with }~~~
d{\vec r}{~^2}\equiv\sum_k(dx^k)^2.
\ee
$a$ and $b$ are functions of $r$ with $a(\in)=b(\in)=1$. The background metric, is in
Minkowski coordinates,
\be
\Bar {ds}^2=  (d{x^0})^2- d\vec r{~^2}
\ee
The   components of $\hat g^{\mu\nu}$ and the non-zero $L$'s   are as
follows
\ba
\hat g^{00}&=&\fr{b^3}{a}~~~{\rm and}~~~\hat g^{kl} =-ab\de^{kl}.~~~{\rm
Moreover~ with}~~~n^k\equiv\fr{x^k}{r},
\\ L_{k00}&=&a^2\lll  3  \fr{b'}{b}-\fr{a'}{a} \rrr n^k ~~~,~~~L_{kmn}=-b^2\lll    
\fr{b'}{b}+\fr{a'}{a} \rrr n^k\de_{mn}~~~{\rm and}\\
 L^{k00}&=&-\fr{1}{a^2b^2}\lll  3  \fr{b'}{b}-\fr{a'}{a} \rrr n^k~~~,~~~
L^{kmn}=\fr{1}{b^4}\lll    \fr{b'}{b}+\fr{a'}{a} \rrr n^k\de^{mn}.
\ea
With these expressions one  can  calculate  the gravitational energy:
\be
E_G=-\1k\int_\infty  {\lll b^{-1}\rrr' }^2dV~~~{\rm where}~~~dV=b^3d^3x
\ee
Here and later on, $dV$ represents the proper volume element.  This simple
formula shows that
$E_G<0$. Moreover it gives a nice 3-coordinate independent
gravitational energy density:
\be
\ep_G\equiv=-\1k {\lll b^{-1}\rrr' }^2
\label{37}
\ee
Using the \Sc solution for empty space   in isotropic coordinates, we also find
  that
\be
E_G=-\1k\int_{r\le r_M}{\lll b^{-1}\rrr' }^2dV-\ha\fr{GM^2}{r_M}
\label{38}
\ee
$r_M$ is the  radial isotropic coordinate  boundary of the   matter. If the
matter is    a thin hollow shell the integral is zero.

By definition $E_G$ is the same as the $\Om$ of MTW \cite{MTW}. To
make the connection explicit consider the same metric in \Sc coordinates  
\be
ds^2=a^2 dt^2-B^2dR^2-R^2d\om^2 ~~~{\rm where}~~~b\fr{dr}{dR}=B
~~~{\rm and}~~~ br= R.
\label{39}
\ee
Consider also one of Einstein's \eeee in isotropic coordinates $G^0_0=\ka T^0_0$
or, see \cite{To}:
\be
 -\fr{4\sqrt{b}}{b^3r^2}\LLL r^2(\sqrt{b})'  \RRR'=\ka\rho c^2.
\label{310}
\ee
One easily find, using (\ref{39}),  (\ref{310}) and the well known
solution\footnote{which we shall only need    for $r\ra\in$.} of  (\ref{310}) for
$\rho=0$ i.e.
\be
\sqrt{b}=1+\fr{GM/c^2}{2r},
\label{311}
\ee
that (\ref{38}) can be rewritten in the form given in \cite {MTW} which in our
notation reads:
\be
E_G=-\int\lll 1-B^{-1} \rrr \rho c^2dV.
\ee  


 \sect{ $\bf{  E_G}$ for conformastationary metrics}
The gravitational energy can also be explicitly evaluated for
the far less symmetric conformastationary metrics \cite{ES}  which may be
written as follows \cite{KBL99}:
\be
ds^2=f(dx^0-\AAA_kdx^k)^2-f^{-1}d\vec r^{~2}.
\ee
It has been noticed in \cite{BK} that this metric is     in harmonic
coordinates provided, here at least, that
$\AAA_k$ is divergenceless. This leaves no gauge freedom except for global
translations and Lorentz rotations.

Solutions of Einstein-Maxwell's equations with this form of metric have been the
object of much scrutiny in empty\footnote{ See \cite{ES} for a revue of
the subject.} and non-empty spaces \cite{KBL99}. 
The \bbb metric is again $c^2dt^2-d\vec r^{~2}$. 
The elements we need to
calculate $E_G$ are $\hat g^{\mu\nu}$, $w^\mu$ and $L_\rho^{~\mu\nu}$. First
we write $\gmn$ and $\hat g^{\mu\nu}$: 
\ba
g_{00}&=&f~~,~~g_{0l}=-f\AAA_l~~,~~g_{kl}=-f^{-1}\de_{kl}+f\AAA_k
\AAA_l~~,~~\sqrt{-g}=f^{-1},\\
\hat g^{00}&=&f^{-2}-|\vec\AAA|^2~~,~~\hat g^{0l}=-\AAA_l~~,~~\hat
g^{kl}=-\de^{kl}~~{\rm with}~~|\vec\AAA|^2\equiv\sum_k\AAA_k^2.
\ea
Next, following (\ref{218}),  
\be
w^0=\fr{1}{\sqrt f}~~~,~~~w^k=0.
\ee
 We shall considerably simplify the formulas for the $L$'s by using  where
possible a vector notation for 3-vectors {\it with indices down}\footnote{with
one exception see (\ref{313}).} like
$\vec\AAA=\{\AAA_k\}$. Thus the non zero $L$'s are
\ba
L^{000}&=&-f^2\vec\AAA\cdot\vec\na(f^{-2}-|\vec\AAA|^2)~~,~~L^{k00}=-f^2
\di_k(f^{-2}-|\vec\AAA|^2),\\
L^{00l}&=&f^2\vec\AAA\cdot\vec\na\AAA_l~~,~~L^{kl0}=f^2\di_k\AAA_l,\\
L_{k00}&=&-2\di_kf~~,~~L_{kl0}=f^{-1}\di_k(f^2\AAA_l)~~,~~L_{kmn}
=-f^{-1}\di_k(f^2\AAA_m\AAA_n).
\ea
With these expressions one easily calculate $E_G$ which, as we shall see  is  
negative definite. Set
\be
E_G=\int_{\in }\ep_G dV ~~~{\rm with}~~~dV=\lll  1-f^2|\vec\AAA|^2\rrr^\ha
f^{-\fr{3}{2}}d^3x. 
\ee
Then the gravitational energy density $\ep_G$, a scalar in 3-space, is
\be
\ep_G=-\fr{1}{4\ka} \LLL f^{-1}|\vec\na
f|^2+f^3|\vec\na\times\vec\AAA|^2 +f^2\vec\na f\cdot
\vec\AAA\times(\vec\na\times\vec\AAA)  
\RRR \lll  1-f^2|\vec\AAA|^2\rrr^{-\ha}. 
\label{49}
\ee
To gain some insight into this formula let us consider a few special cases.
\vs
 {\it (i) Static weak field  }
\vs
This is not a particularly interesting case except to show that in the weak field
approximation we recover the classical Newton G-energy. Indeed in this case
$\vec\AAA=0$ and, setting   
$f\simeq 1+2\phi/c^2$, we have
\be
\ep_G=-\fr{1}{8\pi G} |\vec\na\phi|^2.
\ee
  $\phi$ is the Newtonian potential. This formula can also be recovered from
(\ref{37}) with (\ref{311}).
 
\vs
 {\it (ii) Static   fields  }
\vs
If $\vec\AAA=0$ but $f$ is not nearly equal to one, the metric is that of
Papapetrou
\cite{Pa} and Majumbar \cite{Ma} and 
\be
\ep_G=-\fr{1}{8\pi G}  |\vec\EE|^2 ~~{\rm
where}~~\vec\EE\equiv-c^2\vec\na\ln\sqrt{f}.
\ee
 $\vec\EE$
acts as a potential force or what is also referred to as a ``gravoelectric" force
per unit mass    on test particles (see below). Synge
\cite{Sy}, who called those metrics  ``conformastatic", showed that   
solutions of Einstein's equations of this form appear for static charged dust with electric
and gravitational forces in exact balance; the electric charge density
$\si=\sqrt{G}\rho  $. It is interesting to note that the electric field in this
case\footnote{See for instance
\cite{LBK}. Incidentally on page 8 of that paper there is a printing mistake in the
expression for $\vec E$: there must be a $1/\sqrt{G}$.} is
$\vec E=\fr{c^2}{\sqrt{G}}\vec\na
\sqrt{f}$ and the corresponding electromagnetic energy density
\be
\ep_{EM}=\fr{1}{8\pi} |\vec E|^2 =-\ep_G.
\ee
There is no local force in the matter. There  is also no local field energy
density in this case. The total energy is equal to the mass-energy of the dust. All
this seems to make good sense. 
\vs
 {\it (iii) Non static   fields  }
\vs
In the general case, $\ep_G$ can be written in terms of the gravoelectric field
$\vec\EE$ and a ``gravomagnetic" field 
\be
\vec\BB=c^2\sqrt{f}\vec\na\times\vec\AAA.
\ee
The reason for this electromagnetic analogy is that the equation for slow
motions of test particles, in strong fields {\it as seen in   a
3-space\footnote{This is the 3-space in which proper lengths are measured.}
with metric
$ f^{-1}\de_{kl}$}
\cite{LL},  with velocity $\vec v$ and momentum $\vec p$,
\be
\vec v=\{v^k=c\fr{dx^k}{ds}\equiv \fr{dx^k}{d\tau}\}~~~{\rm and~
 }~~~\vec p=\{p_k= f^{-1}m  v^k\}~~~{\rm is}~~~\fr{d\vec p}{d\tau}=m\lll
\vec\EE+\fr{\vec v}{c}\times \vec \BB   \rrr.
\label{313}
\ee 
In terms of $\vec\EE$ , $\vec \BB $ and $\vec\AAA$, expression (\ref{49}) can be
written
\be
\ep_G=-\fr{1}{8\pi G}  \LLL   f(1-\fr{1}{4}f^2|\vec\AAA|^2) |\vec\EE|^2+
\fr{1}{4}|\vec\BB+\sqrt{f}\vec\AAA\times\vec\EE|^2+
\fr{1}{4}f^3(\vec\AAA\cdot\vec\EE)^2
\RRR (1-f^2|\vec\AAA|^2)^{-1/2}.
\ee
One can see from this expression that $\ep_G$ is manifestly negative. 
 For
comparison we notice that electromagnetic energy density in this spacetime
has a similar form. Denoting by  $ \vec E=\{E_k\equiv F_{k0}=\di_kA_0\}$ and 
$ \vec B=\{B_k\equiv \sum \ep_{kmn}\di_mA_n\}$ the electromagnetic vector
fields:
\be
\ep_{EM}=\fr{1}{8\pi} \LLL  (1-f^2|\vec\AAA|^2)|\vec E|^2+f^2|\vec
B|^2 + f^2(\vec\AAA\cdot\vec E)^2  
\RRR (1-f^2|\vec\AAA|^2)^{-1/2}.
\ee


 \sect{ $\bf{  E_G}$ of a weak stationary field}

In Minkowski coordinates in the background, the metric of the foreground is
usually written
 \be
ds^2=(\eta_{\mu\nu}+h_{\mu\nu})dx^\mu dx^\nu,
\ee
 $\eta_{\mu\nu}=\rm diag ~({1,-1,-1,-1})$ and $|h_{\mu\nu}| \ll 1$. Then,
 \be
\hat
g^{\mu\nu}=\eta^{\mu\nu}-(h^{\mu\nu}-
\ha\eta^{\mu\nu}h)\equiv\eta^{\mu\nu}-\td h^{\mu\nu}~~~{\rm with}
~~~h=\eta^{\rho\si}h_{\rho\si}~~~{\rm
and}~~~L_\rho^{~~\mu\nu}=-\di_\rho\td h^{\mu\nu}.
\ee 
In harmonic coordinate,  $\di_\nu\td h^{\mu\nu}=0$, there is no
further gauge freedom   available except  uniform translations and Lorentz
rotations. Stationarity implies 
$\di_0\td h^{\mu\nu}=0$ while Einstein's equations become
\be
\na^2\td h^\mu_\nu=2\ka T^\mu_\nu.
\label{34} 
\ee
Static observers have velocity components
\be
w^0=1-\ha h^0_0 ~~~{\rm and}~~~w^k=0.
\ee
 If we regard the  $h_k^0$ as components of a vector $\vec h$ in the
background whose curl   $(\vec\na\times\vec h)=\{\sum\ep_{mkl}\di_k h^0_l\}$,
and further decompose $\td h^m_n  $ into a traceless part $_T\td h^m_n  $  and a
trace  $\td h^m_m  $, we find that
\be
\ep_G=-\fr{1}{4\ka} |\vec\na\times\vec
h|^2 -\fr{1}{16\ka} \lll 
 |\vec\na \td h^0_0|^2 +\th |\vec\na \td h^k_k|^2 -2 
 {\vec\na} _T{\td h^m_n}
\cdot \vec\na _T{\td h^n_m}\rrr\nn\\
 ~{\rm where}~ _T{\td h^m_n}\eq\td h^m_n-\th\de^m_n \td h^k_k.
\label{36}
\ee
Note that the expression in terms of $h$'s instead of $\td
h$'s is exactly the same.
 We see that $\ep_G$  is not manifestly negative. However, if  $\vec\na _T{\td
h^m_n}=0$ or is negligible, then G-energy is indeed negative.  This is, for instance,
the case if the source of gravity is  a perfect fluid  for which
$T^0_0\equiv\rho c^2$ and
$T^m_n\equiv-P\de^m_n$. In this case, Einstein's equations (\ref{34}) imply that
$\na^2( _T{\td h^m_n})=0$ because   $T^m_n$ is now traceless. From this follows,
with usual boundary and regularity conditions, that $\vec\na _T{\td h^m_n}=0$ and,
\be
\ep_G=-\fr{1}{4\ka} |\vec\na\times\vec
h|^2-\fr{1}{16\ka} \lll 
 |\vec\na \td h^0_0|^2 +\th |\vec\na \td h^k_k|^2  \rrr< 0.
\ee
The equation of slow motion of a test particle in the weak field has the form
(\ref{313}) with
\be
\vec\EE=-c^2\vec\na\fr{h^0_0}{2}~~~{\rm
and}~~~\vec\BB=c^2\vec\na\times\vec h.
\ee
 If $|\vec\na \td h^k_k|\ll |\vec\na \td h^0_0|$, like in a
perfect fluid  with non relativistic pressure or tension, $| P| \ll\rho c^2$, the
gravitational energy density is:  
\be
\ep_G=-\fr{1}{8\pi G} 
\lll|\vec\EE|^2 +\fr{1}{4}|\vec\BB |^2\rrr.
\ee
 The source of $E_G$ may be a fluid in fast, but not too fast and not necessarily
uniform  rotational  
  motion,  with vorticity, for instance, like in a Dedekind ellipsoid.

Notice that Landau and Lifshitz \cite{LL}  prefer to regard the gravomagnetic
force as a Coriolis force   produced in an orthogonal  frame rotating with
absolute angular velocity  $\vec \Om=\fr{c}{2}\vec\na\times\vec h$.  


 \sect{Comments}

In a similar vein we can calculate gravitational energy of cosmological
perturbations on a Robertson-Walker background. These \bbbb all have a
conformal timelike Killing field. Thus one may be facing  ``conformal
matter-energy" and ``conformal G-energy". Conservation laws and
additivity   of different forms of energy remains valuable knowledge but
the conserved vector  may  not have the simple form it has   here  in
(\ref{217}) and the interpretation of globally conserved quantities is quite
different from asymptotically flat spacetimes.

It is possible that $E_G$ is negative for any stationary  metric with reasonable
physical sources, that is, those whose \em tensor satisfy reasonable energy
conditions.  It is, however,  interesting to notice that $E_G$ is negative in some \ssss
with special symmetries irrespective of any energy condition.
 
\vs
 {\bf AKNOWLEDGMENTS}
\nl\nl
I am much indebted to Alexander Petrov and Nathalie Deruelle for interesting remarks on a
first draft of this work. I am also particularly thankful to Donald Lynden-Bell and J\v\i
r\'\i~Bi\v c\'ak with whom I enjoyed discussing the subject all along and from
whom I received much helpful advice.

\vs

\vs

  \end{document}